\documentclass{osa-article}

\journal{osajournal}
\articletype{Research Article}

\begin{document}

\title{Nanoscale compositional evolution in complex oxide based resistive memories}

\author{Taimur Ahmed,\authormark{1,*} Sumeet Walia,\authormark{1} Edwin L. H. Mayes,\authormark{2} Rajesh Ramanathan,\authormark{3} Paul Guagliardo,\authormark{4} Vipul Bansal,\authormark{3} Madhu Bhaskaran,\authormark{1} J. Joshua Yang,\authormark{5} and Sharath Sriram\authormark{1}}

\address{\authormark{1}Functional Materials and Microsystems Research Group and Micro Nano Research Facility, RMIT University, Melbourne, VIC 3001, Australia\\
\authormark{2}RMIT Microscopy and Microanalysis Facility, RMIT University, Melbourne, VIC 3001, Australia\\
\authormark{3}Ian Potter NanoBioSensing Facility, NanoBiotechnology Research Laboratory, School of Science, RMIT University, Melbourne, VIC 3001, Australia\\
\authormark{4}Centre for Microscopy, Characterisation and Analysis, The University of Western Australia, Perth, WA 6009, Australia\\
\authormark{5}Department of Electrical and Computer Engineering, University of Massachusetts, Amherst, MA 01003, USA}

\email{\authormark{*}taimur.ahmed@rmit.edu.au}

\begin{abstract*}
Functional oxides based resistive memories are recognized as potential candidate for the next-generation high density data storage and neuromorphic applications. Fundamental understanding of the compositional changes in the functional oxides is required to tune the resistive switching characteristics for enhanced memory performance. Herein, we present the micro/nano-structural and compositional changes induced in a resistive oxide memory during resistive switching. Oxygen deficient amorphous chromium doped strontium titanate (Cr:$a$-SrTiO$_{3-x}$) based resistance change memories are fabricated in a Ti/Cr:$a$-SrTiO$_{3-x}$ heterostructure and subjected to different biasing conditions to set memory states. Transmission electron microscope based cross-sectional analyses of the memory devices in different memory states shows that the micro/nano-structural changes in amorphous complex oxide and associated redox processes define the resistive switching behavior. These experimental results provide insights and supporting material for Ref.\cite{CrSTO}.
\end{abstract*}

\section*{Introduction}
In recent years, a variety of materials exhibiting resistive switching behavior, such as binary metal oxides and transition metal oxides, have been reported. Strontium titanate, a model perovskite oxide system, has shown significant potential for valance change resistive switching memories owing to its tunable defect chemistry via doping, mixed anionic-electronic and non-linear switching dynamics \cite{Nili,Tech,arx2018,Nano2017}. Despite enormous characterization efforts, the driving nanoscale redox processes and associated micro/nano compositional changes in the oxide system are still not completely clear. Here, we utilize different cross-sectional transmission electron microscope (TEM) analyses techniques, such as energy dispersive x-ray spectroscopy (EDS) and electron energy loss spectroscopy (EELS), to study the effect of biasing conditions and device structure on the resistive switching behavior in Ti/Cr:$a$-STO$_{x}$ based memory cells. The micro-structural analyses shows that top Ti layer introduces a Ti$_{2}$O$_{3}$  interfacial layer with Cr:$a$-STO$_{x}$ and acts as an oxygen reservoir during redox processes. Furthermore, formation of oxygen deficient filamentary paths at locally crystalline regions of Cr:$a$-STO$_{x}$ control the resistive switching behavior. These results provide supporting material for our recent report on inducing different resistive switching behaviors in Cr:$a$-STO$_{x}$ based resistive memories \cite{CrSTO}. 

\section*{Results}
The memory devices are fabricated in metal-insulator-metal (MIM) configuration by following the fabrication steps in Ref.\cite{CrSTO} on SiO$_{2}$/Si substrates. In the MIM structure Pt (35 nm)/Ti (8 nm) serves as a top metal electrode, Cr:$a$-STO$_{x}$ (25 nm) as an insulator and Pt (7 nm)/Ti (3 nm) as a bottom metal electrode. In order to ascertain the effect of applied bias (during electroforming and resistive switching) on the metal/oxide interfaces and within the functional oxide (Cr:$a$-STO$_{x}$), the cross-sectional TEM lamellae are prepared \textit{via} focused ion beam (FIB) cuts from separate MIM memory cells subjected to different biasing conditions namely; pristine, electroformed and switching devices.
The cross-sectional TEM and EDS analyses of a pristine Cr:$a$-STO$_{x}$ MIM device is presented in Fig 1. The TEM micrograph (Fig. 1a) of the MIM structure shows the Cr:$a$-STO$_{x}$ oxide film is sandwiched between the top Pt/Ti and bottom Pt electrodes. The selected-area electron diffraction (SAED) pattern (Fig. 1b) and high-resolution TEM (HRTEM) micrograph (Fig. 1c), collected from the MIM cross-section show an amorphous structure of the Cr:$a$-STO$_{x}$ oxide film. Fig. 1d shows a high-angle annular dark field (HAADF) micrograph of the pristine cross-section in scanning TEM. Fig.1(e$-$i) present elemental EDS maps of Pt, Sr, Ti, Cr and O, respectively. The EDS maps confirm successful Cr doping in the $a$-STO$_{x}$ (\textit{via} co-sputtering of Cr and $a$-STO$_{3}$) and also the desired MIM structure of the memristive devices to execute multiple resistive switching behaviors. 
\paragraph*{}
The electronic composition and the relative distribution of oxygen content across the pristine MIM device is presented in the EELS area map (Fig. 2a), collected by considering the O$-$\textit{K} edge intensities. The EELS Ti$-$\textit{L}$_{2,3}$ and O$-$\textit{K} edge profiles (Fig. 2b) are also obtained from a line-scan across the MIM structure. Broad Ti$-$\textit{L}$_{3}$ and Ti$-$L$_{2}$ peaks at the top Ti/Cr:$a$-STO$_{x}$ interface can be used to identify the oxidation states of Ti (\textit{i.e}., presence of mixed Ti$^{2+}$ and Ti$^{3+}$  oxidation states at the top interface) as explained in Ref \cite{TEM}.
\paragraph*{}
The electroforming of Cr:$a$-STO$_{x}$ MIM devices is carried out by applying a negative voltage ($<$-5 V) on the bottom Pt electrode to exhibit clockwise bipolar (CW-BP) resistive switching behavior, prior to the TEM lamella preparation. The highlighted region of interest (ROI, enclosed in a box in Fig. 3a) presents an example of isolated incomplete filaments along the bottom interface. The fast Fourier transform (FFT, Fig. 3b) taken from the ROI presents the diffraction spots. The diffraction spots indicated with arrows can be indexed to the cubic phase of STO and are masked to perform an inverse FFT (iFFT), as shown in Fig. 3c. Several diffraction spots are also used to ensure the iFFT show the distribution of crystals with differing orientations.
\paragraph*{}
Fig. 4 presents the cross-sectional HR-TEM micrograph of the top Ti/Cr:$a$-STO$_{x}$ interface of a MIM device subjected to the electroforming. In order to assess the morphological changes in the top Ti layer, two ROIs are selected along its length at two different locations (Fig. 4a). The FFT diffraction patterns generated for each location (Fig. 4b,c) show the polycrystalline structure of the top Ti layer. The high intensity diffraction patterns with the \textit{d}-spacing, ranging between 2.4 $-$ 2.6 \AA,  can be indexed to the different planes of the rhombohedral Ti$_{2}$O$_{3}$.
\paragraph*{}
Fig. 5a shows the cross sectional HRTEM micrograph of a Cr:$a$-STO$_{x}$ MIM device in its HRS, exhibiting CW-BP resistive switching characteristics. The ROI shows the localized crystalline region in the active Cr:$a$-STO$_{x}$ layer, extending from the bottom Pt electrode. The FFT diffraction pattern of the ROI (Fig. 5b) shows the presence of different crystalline phases of STO. The diffraction spot of the highest intensity (marked as spot 1 in Fig. 5b) and other weaker diffraction spots can be assigned to the cubic perovskite STO phase. However, the encircled diffraction spots could not be assigned to the cubic perovskite STO phase. The spot 1 (with the \textit{d}-spacing of 2.8 \AA) is used to generate the iFFT (Fig. 5c) highlighting the presence of [011] cubic STO phase in the selected ROI.
\paragraph*{}
Fig. 6 shows the EELS O$-$\textit{K} edge area map of the locally crystalline ROI from the MIM device exhibiting CW-BP resistive switching behavior (presented in Fig. 5). The representative MIM device is set to HRS prior to TEM sample preparation. Relatively low oxygen content at the bottom Pt electrode shows a ruptured filamentary path and accumulation of the V$_{o}$s at anode in HRS. Presence of varying oxygen content at the vicinity of top Pt electrode shows the oxidation of Ti layer to a sub-stoichiometric Ti$_{2}$O$_{3}$, as indexed in the FFT analysis (as presented in Fig. 4).
\section*{Acknowledgments}
The authors acknowledge support from the Australian Research Council (ARC) for personnel and project support via DP130100062 (S.S.), DE160100023 (M.B.), and FT140101285 (V.B.) and equipment funding through LE0882246, LE0989615, LE110100223, and LE150100001. The authors would like to acknowledge the technical assistance of the Micro Nano Research Facility (MNRF) and the RMIT Microscopy and Microanalysis Research Facility (RMMF). Also Australian Microscopy and Microanalysis Research Facility, AuScope, the Science and Industry Endowment Fund, and the State Government of Western Australian for contributing to the Ion Probe Facility at the Centre for Microscopy, Characterisation and Analysis at the University of Western Australia.

\bibliography{TAhmed_CrSTO_Lib}

\begin{figure}[h!]
\centering\includegraphics[width=10cm]{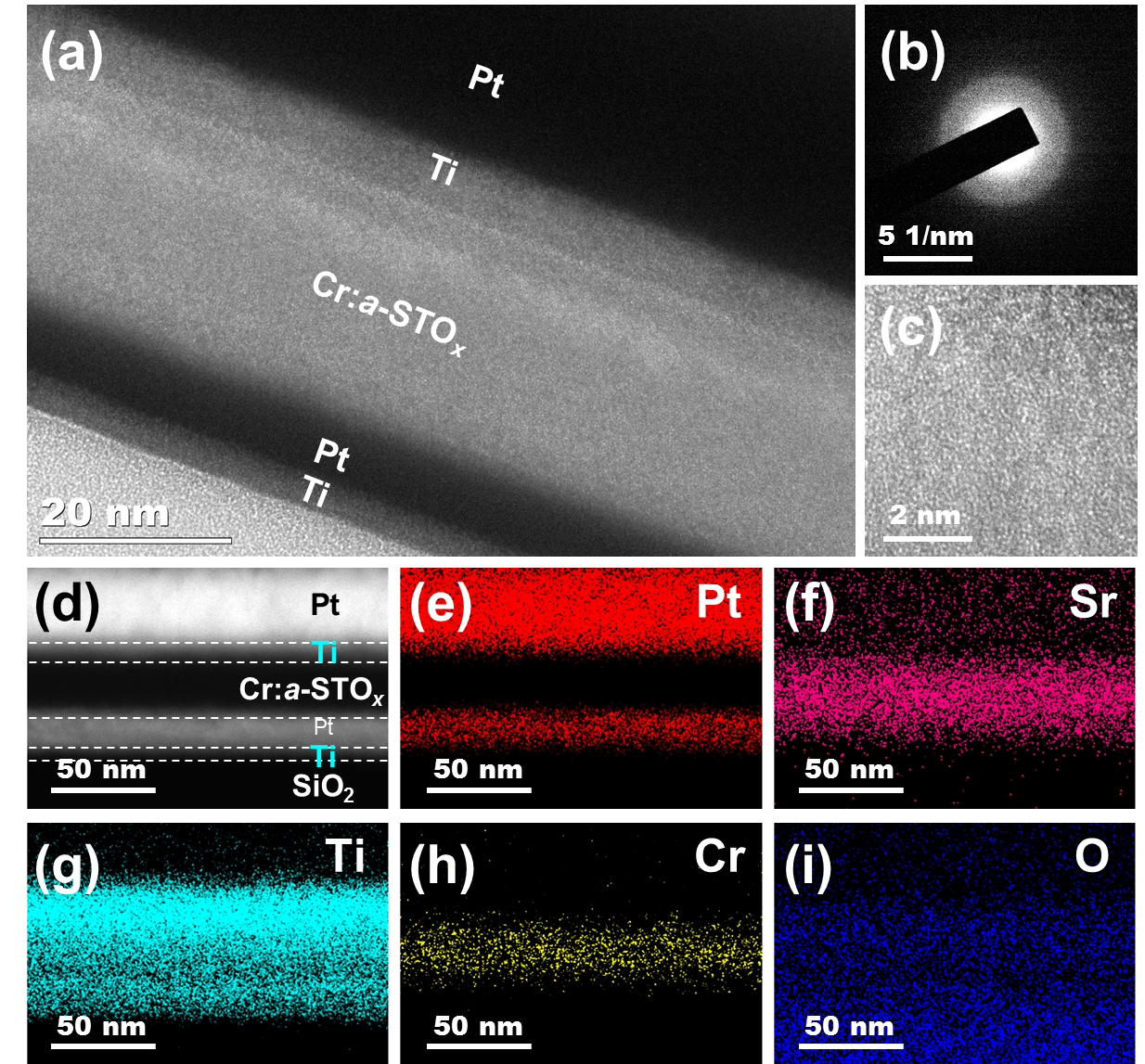}
\caption{Microstructure structure of the pristine Cr:$a$-STO$_{x}$ devices. (a) TEM micrograph of a pristine MIM device. (b) Selected area electron diffraction pattern collected from the MIM cross-section. (c) High resolution TEM micrograph of the Cr:$a$-STO$_{x}$ oxide film. (d) High-angle annular dark field TEM micrograph of the pristine MIM device.The elemental EDS mapping of (e) Pt, (f) Sr, (g) Ti, (h) Cr and (i) O.}
\end{figure}

\begin{figure}[!h]
\centering\includegraphics[width=11cm]{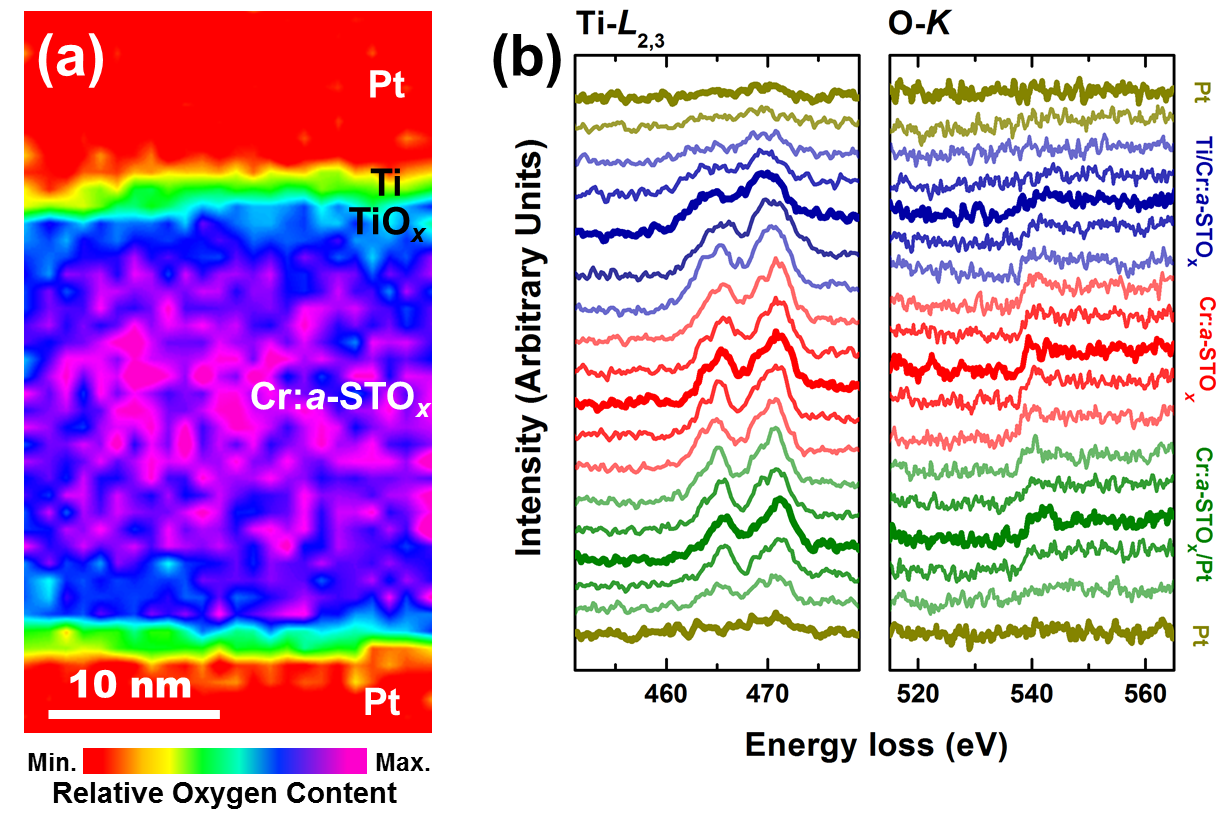}
\caption{Electronic structure of the pristine Cr:$a$-STO$_{x}$ devices. (a) The EELS O$-$\textit{K} edge area map and (b) the EELS Ti$-$\textit{L}$_{2,3}$ and O$-$\textit{K} edge profiles along a line scan across the MIM device}
\end{figure}

\begin{figure}[h!]
\centering\includegraphics[width=9cm]{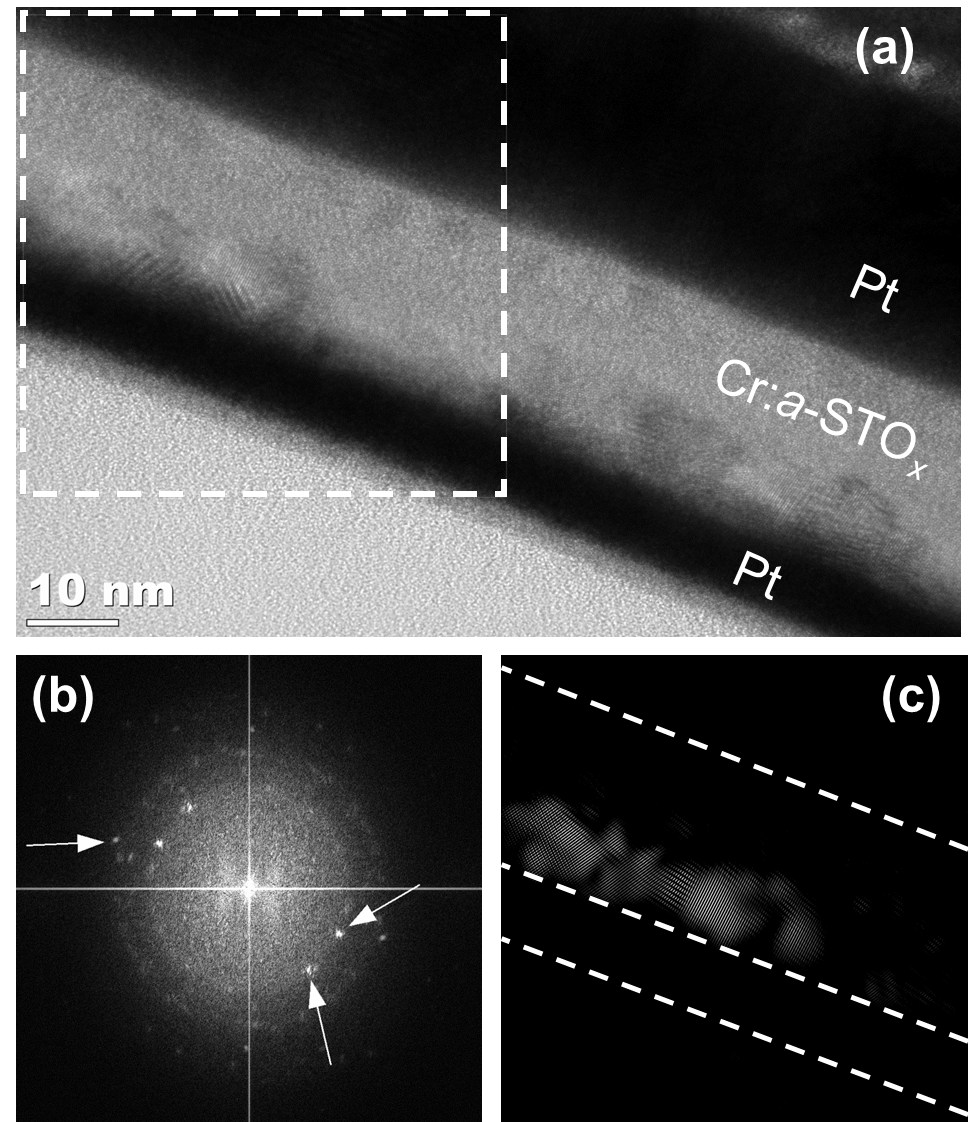}
\caption{Morphological analyses of electroformed devices. (a) TEM micrograph of the MIM device subjected to electroforming step. The box encloses the ROI. (b) The FFT diffraction patterns generated from the ROI enclosed in (a). (c) The iFFT obtained from a diffraction spot in (b) highlight the crystalline region along the bottom electrode.}
\end{figure}
\begin{figure}[ht!]
\centering\includegraphics[width=13cm]{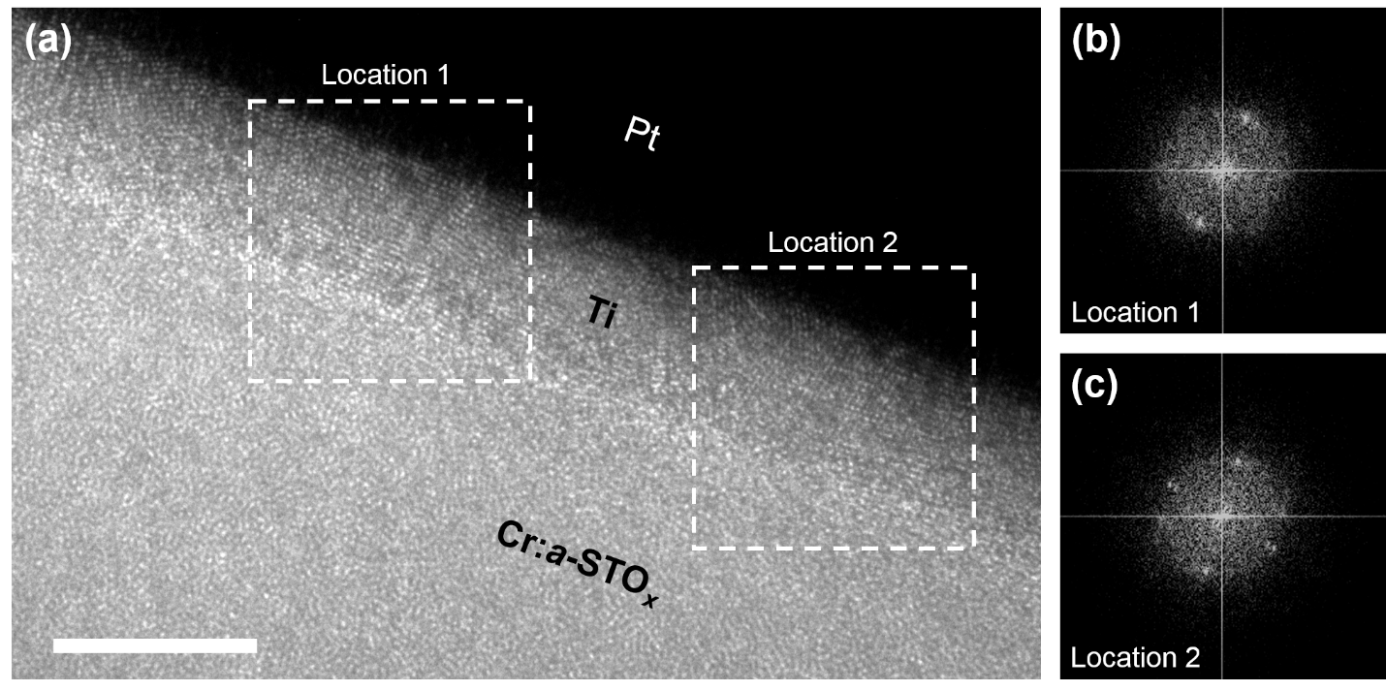}
\caption{Micro-structure of top Pt/Ti/Cr:$a$-STO$_{x}$ interface of electroformed MIM devices. (a) TEM micrograph of top interface. Two ROIs are selected at Location 1 and Location 2, enclosed in boxes. Scale bar denotes 10 nm. (b) and (c) are the FFT diffraction patterns generated from Location 1 and Location 2, respectively, in (a).}
\end{figure}
\begin{figure}[h!]
\centering\includegraphics[width=10cm]{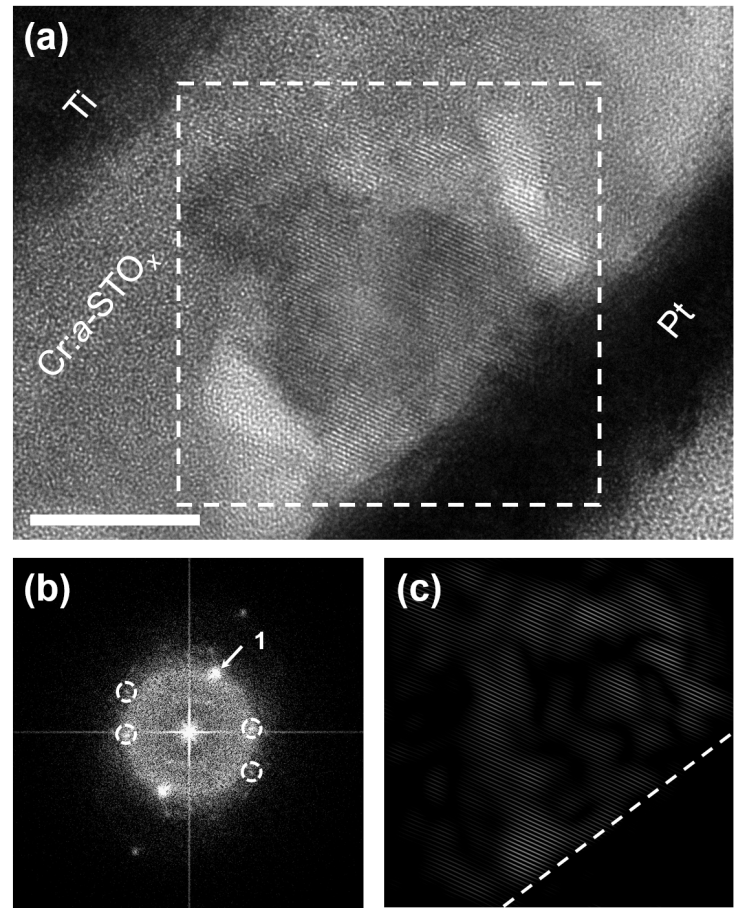}
\caption{Morphological analyses of the Cr:$a$-STO$_{x}$ MIM devices in HRS and exhibiting CW-BP resistive switching behavior. (a) TEM micrograph of the MIM device subjected to at least 100 resistive switching cycles and set to HRS prior to the lamella preparation. ROI is enclosed in the box. Scale bars 5 nm. (b) The FFT diffraction patterns generated from the ROI enclosed in (a). (c) The iFFT obtained from spot 1 in (b) highlight the crystalline region}
\end{figure}
\begin{figure}[h!]
\centering\includegraphics[width=7cm]{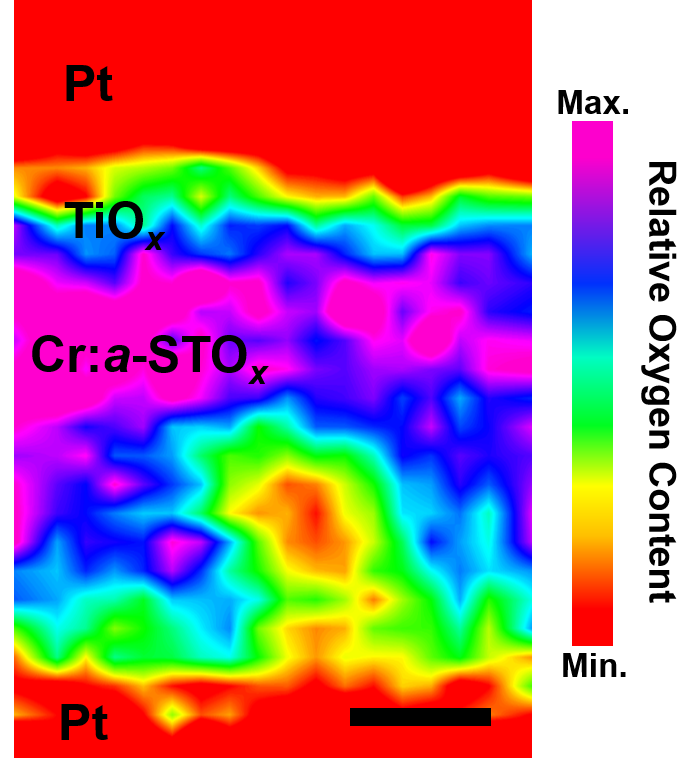}
\caption{The EELS O$-$\textit{K} edge area map of the raptured conductive filamentary path in HRS. Scale bar represents 20 nm.}
\end{figure}

\end{document}